\documentclass[epj]{svjour}
\usepackage{graphicx}
\usepackage{amsmath,amssymb}
\usepackage{color}

\newcommand{\nin}{\noindent}

\begin{document}
\sloppy

\title{Non-adiabatic extension of the Zak phase and charge pumping in the Rice--Mele model}

\author{Yoshihito Kuno}

\institute{Department of Physics, Graduate School of Science, Kyoto University, Kyoto 606-8502, Japan}
	
\date{Received:  / Revised version: }

\abstract{
In this study, the Landau--Zener (LZ) transition method is applied to investigate a weak non-adiabatic effect on the Zak phase and the topological charge pumping in the Rice--Mele model. 
The non-adiabatic effect is formulated using the LZ transfer matrix. 
The effective lower band wave function picks up the Stokes phase as well as the usual dynamical phase through two avoided crossings 
appearing in the two band instantaneous energy spectrum. 
The interference effect from the upper band has a decisive influence on the decay behavior of the lower band population.
A non-adiabatic extension of the Zak phase can then be formulated, corresponding to the center of mass of the lower band Wannier function. 
Furthermore, we estimate the validity of the LZ formalism and verify the breakdown of the quantization of the topological charge pumping by changing the sweeping speed.   
}


\maketitle	
\section{Introduction}
\label{sec:Intro}

The theory of topological physics has been realized and is being investigated in detail using real experimental systems. 
Specifically, systems of cold atoms in optical lattices have a significant possibility of simulating the physics because such systems have a high parameter controllability, 
isolation from environment, and no impurities \cite{Bloch,Georgescu}. 
Very recently, as a typical verification experiment in one-dimensional (1D) topological physics, 
topological charge pumping phenomena \cite{Thouless,Shen,Asboth} have been realized in cold atoms in a 1D optical lattice \cite{LeiWang,Lohse,Nakajima}. 
It is therefore important to theoretically consider the topological physics and obtain new knowledge that has yet to be obtained.

Motivated by the experimental successes of the topological charge pumping, various studies on 1D topological physics have been conducted in recent years. 
For example, the interaction effect for the topological charge pumping under adiabatic conditions has been extensively studied \cite{Qian,Zeng,Nakagawa,Tangpanitanon,Kuno,Hayward}. 
The breakdown of the quantization of topological charge pumping has also been discussed \cite{RLi,Privitera}. 
The effect of the initial nonequilibrium state for topological charge pumping and 
interband coherence correction during adiabatic pumping in a periodically driven system were also reported \cite{Zhou,Wang,Raghava}.
The generalization of the Zak phase to the thermal states \cite{Viyuela} and the reinterpretation of the microscopic meaning of the Zak phase \cite{Rhim} have also been reported. 
However, some areas in this field have yet to be investigated in detail. 

During the past three decades, numerous papers on theoretical topological physics have been submitted. 
The fundamental framework of topological physics has been theoretically developed \cite{Shen,Asboth}. 
In conventional topological insulators, the topological properties are based on the following assumptions: the bulk band gap exists, and the system is close to equilibrium, i.e., the model is under adiabatic conditions. 
This naturally brings up a question of how the non-adiabaticity affects the topological properties, 
which needs to be answered.
However, there have been few studies on the theoretical formulation and quantitative evaluation of non-adiabatic effects \cite{Privitera}.  
Therefore, this paper discusses the non-adiabatic effects on the topological properties using a typical model, focusing on the properties of the lower band ground state, particularly the lower band topological properties of the system. The target model is the Rice--Mele (RM) model \cite{Rice}. 
If we introduce an adiabatic modulation parameter in this model and change the model parameter, 
it exhibits a quantization of the Zak phase \cite{Zak} under an inversion symmetric condition. 
In addition, under a certain periodic adiabatic parameter sweeping, 
the model exhibits a two-dimensional instantaneous energy band topology in the adiabatic parameter dimension. 
The instantaneous energy band topology then leads the topological charge pumping. 
The RM model is much close to some experimental cold atom systems in an optical superlattice
\cite{Lohse,Nakajima,Atala}. In this study, we primarily deal with a weak non-adiabatic effect, where the transition probability between the lower and upper bands is small.  
The non-adiabatic dynamics for the lower band population is formulated by applying the Landau-Zener (LZ) transition method \cite{Landau,Zener,Nori}. 
Using the analytical population dynamics, we can formally construct the center of mass (CM) shift for the lower-band Wannier function after one pumping cycle and formulate a lower band pumped charge related to the topological charge pumping.
A similar prescription is described in~\cite{Oka,Lim}. 
After formulating the non-adiabatic effect, the breakdown of the quantization of the CM shift is numerically estimated using the obtained formula.

This paper is organized as follows. In Sec.~\ref{Rice-mele}, the RM model is introduced and the proper linearized form of the RM model for the LZ transition method is explained. 
In Sec.~\ref{LZTheory}, we show the LZ transition method. In particular, the LZ transfer matrix is introduced.
In Sec.~\ref{TEA}, we describe the dynamics of the lower band population under the assumption of an adiabatic-impulse approximation.
In Sec.~\ref{NAEZ}, we formulate the non-adiabatic extension form of the Zak phase and the CM shift of the lower band Wannier function corresponding to an electric polarization \cite{Resta,Vanderbilt,Marzari}. In Sec.~\ref{ELZ},  we estimate the validity of the LZ formulation by using a numerical simulation. 
In Sec.~\ref{EXPLZ}, we discuss the current experimental conditions used for testing our results. 
Finally, the conclusion is given in Sec.~\ref{Conq}.

The aim of this work is different from that of previous works such as \cite{Lim,Lim2,XShen}, 
which focused on the effects of an external force for the system and the Stueckelberg interferometry.

\section{Rice-Mele model}
\label{Rice-mele}
\begin{figure}[t]
\begin{center} 
\includegraphics[width=8.5cm]{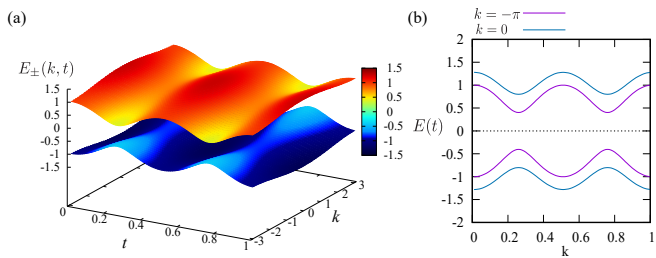}
\end{center} 
\caption{(a) The band structure of the $t-k$ parameter space. (b) The band structure for $k=\pi$ and $0$. Two avoided crossings appear. For both cases, $J_{0}/\Delta_{0}=0.4$ and $\delta_{0}/\Delta_{0}=0.2$}.
\label{RMband}
\end{figure}
We start by considering the bulk momentum representation of the RM model. The Bloch vector representation is written in the following form:
\begin{eqnarray}
\hat{h}_{RM}(k,t)=d_{x}(k,t)\hat{\sigma}_{x}+d_{y}(k,t)\hat{\sigma}_{y}+d_{z}(k,t)\hat{\sigma}_{z},
\end{eqnarray}
where $\hat{\sigma}_{i}$ ($i=x,y,z$) is the Pauli matrix; $d_{x}(k,t)=J_{1}(t)+J_{2}(t)\cos k$; $d_{y}(k,t)=J_{2}(t)\sin k$; and $d_{z}(k,t)=-\Delta_{0}\cos (\Omega t)$, 
$J_{1}(t) = J_{0} + \delta_{0}\sin(\Omega t)$; $J_{2}(t)=J_{0}-\delta_{0}\sin(\Omega t)$; and $\Omega=2\pi/T$. In addition, $T$ is the interval time of one cycle in the RM model. $J_{0}>0$, $\delta_{0}>0$, and $\Delta_{0}>0$.
The instantaneous energy spectrum has two bands, given by $E_{\pm}(k,t)=\pm |{\bf d}|$. The typical instantaneous energy spectrum is plotted in Fig.\ref{RMband} (a).
When we focus on a certain wave number $k$, the $t$ dependent instantaneous energy spectrum at $k$ can be regarded 
as a two-level system including some avoided crossings in a certain parameter regime (see Fig.~\ref{RMband} (b)).
Assuming that the two bands never touch each other along one pumping cycle (interval time $T$), and a specific case $\Delta_{0}>2\delta_{0}$ and $J_{0}>\delta_{0}$, 
an avoided crossing appears around $t=T/4\equiv t_{1}$ and $3T/4\equiv t_{2}$ for {\it any fixed $k$}. 
At the avoided crossings, a non-adiabatic transition may occur depending on the sweeping speed, depending on $\Omega$.
The energy landscape is shown in Fig.\ref{RMband} (a). In what follows, the focus is placed on the energy landscape.
Here, the LZ transition around the avoided crossing point for a certain fixed $k$ is considered. 
Around the avoided crossing points $t=t_1$ and $t_2$, $\hat{h}_{RM}$ can be linearized in terms of $t$ as $t=t_{1}\pm\delta t$ and $t=t_{2}\pm\delta t$. 
The linearized Hamiltonian for $\hat{h}_{RM}$ is generally given in the following form:
\begin{eqnarray}
\hat{h}_{RM}(k,\delta t)=A(k)\hat{\sigma}_{x}+B(k)\hat{\sigma}_{y}+\Delta_{0}\Omega\delta t\hat{\sigma}_{z}.
\label{LZH1}
\end{eqnarray}
Here, $A(k)$ and $B(k)$ are $k$-dependent functions (independent of $t$), 
$A(k)=(J_{0}+\delta_{0})+(J_{0}-\delta_{0})\cos k$, $B(k)=(J_{0}-\delta_{0})\sin k$, and 
the $O (\delta t^{2})$ order terms are dropped.
Let us introduce the rotational transformation of the Pauli matrix. In the spin space, for the $i$-axis rotation ($i=1(x),2(y),3(z)$) with a certain angle $\rho$, 
the rotated $j$-component Pauli matrix $\tilde{\sigma}_{j}$ is given by
\begin{eqnarray}
\tilde{\sigma}_{j}(\rho)\equiv \hat{\sigma}_{j}\cos\rho+\epsilon_{ijk}\hat{\sigma}_{k}\sin\rho. 
\end{eqnarray}
Here, we use the above rotational transformation twice: The first is the $z$-axis rotation with $\rho=\tan^{-1}(B(k)/A(k))$, 
and the second is the $y$-axis rotation with $\rho=-\pi$. 
Thus, $\hat{h}_{RM}(k,\delta t)$ can be transformed into the following form: 
\begin{eqnarray}
\tilde{h}_{RM}(k,\delta t)=-\sqrt{A^{2}(k)+B^{2}(k)}\tilde{\sigma}_{x}-\Delta_{0}\Omega\delta t\tilde{\sigma}_{z},
\label{LZH2}
\end{eqnarray}
where $\tilde{\sigma}_{x(z)}$ is a rotated $x$($z$)-component Pauli matrix. In addition, $\tilde{h}_{RM}(k,\delta t)$ is the canonical form for applying the LZ transition formula. 

Now, the general LZ application form is defined as $\tilde{h}_{RM}(k,\delta t) \equiv -\frac{\Delta(k)}{2}\tilde{\sigma}_{x}-\frac{v\delta t}{2}\tilde{\sigma}_{z}$, 
and thus $\Delta (k) \equiv2\sqrt{A^{2}(k)+B^{2}(k)}$ and $v=2\Delta_{0}\Omega$. 
Here, the adiabaticity parameter is introduced by $\bar{\delta}(k)=\Delta^{2}(k)/(4v)$. 
In the LZ transition method, the transition probability for the upper band at an avoided crossing is given as $\exp[-2\pi \bar{\delta}(k)]$. 
Thus, $\bar{\delta}(k)$ characterizes the degree of the adiabaticity. 
In this study, we assume the weak nonadiabatic regime. This means that $\bar{\delta}(k)$ is large to a certain extent, 
that is, the transition probability for the upper band at an avoided crossing is small (Practically, we assume that the transition probability is less than 50\% for all $k$). 
In what follows, we set $\hbar=1$ and take $\Delta_0$ as the unit of energy, $\Delta_0=1$. 

\section{Landau-Zener transition}
\label{LZTheory}
The LZ transition is the transition between the lower and upper bands at avoided crossings.
The linearized RM model $\tilde{h}_{RM}(k,\delta t)$ of Eq.~(\ref{LZH2}) has two avoided crossings, namely, at $t=t_1$ and $t=t_{2}$. 
The linearized Hamiltonian is given in the following matrix form
\begin{eqnarray}
&&\tilde{h}^{\pm}_{RM}(k,\delta t)=\nonumber\\
&&\left[ 
\begin{array}{cc} 
\pm \Delta_{0}\Omega\delta t & \sqrt{A^{2}(k)+B^{2}(k)}e^{i\tilde{\rho}(k)}  \\ 
\sqrt{A^{2}(k)+B^{2}(k)}e^{-i\tilde{\rho}(k)}& \mp \Delta_{0}\Omega\delta t \\ 
\end{array}
\right],
\label{LZH4}
\end{eqnarray}
where $\tilde{h}^{+(-)}_{RM}(k,\delta t)$ is defined around $t=t_{1(2)}$ and $\tilde{\rho}(k)=\tan^{-1}[B(k)/A(k)]$. 
From the above matrix, the LZ transfer matrix can be introduced around the two avoided crossing points at $t=t_{1}$ and $t_{2}$. 
The matrix is known to have the following form: \cite{Nori,Lim,Lim2,Kayanuma,Kayanuma2}: 
\begin{eqnarray}
&&\Gamma_{\pm} (k) =\nonumber\\
&&\left[ 
\begin{array}{cc} 
\sqrt{q_{LZ}(k)}e^{-i(\gamma_{nb}(k)\mp \tilde{\rho}(k))} & \pm \sqrt{p_{LZ}(k)}  \\ 
\mp \sqrt{p_{LZ}(k)} & \sqrt{q_{LZ}(k)}e^{i(\gamma_{nb}(k)\pm \tilde{\rho}(k))}  \\ 
\end{array}
\right],\nonumber\\
\label{LZ_matrix}
\end{eqnarray}
where $\gamma_{nb}(k)$ is the Stokes phase \cite{Kayanuma,Kayanuma2}, 
$p_{LZ}(k)=e^{-2\pi \bar{\delta}(k)}$ is the transition probability between the lower and upper bands, and $q_{LZ}(k)=1-p_{LZ}(k)$.
In addition, $\gamma_{nb}(k)$ is given by the following form:
\begin{eqnarray}
\gamma_{nb}(k)=\frac{\pi}{4}+\bar{\delta}(k)[\ln \bar{\delta}(k)-1]+ {\rm arg}[\Gamma (1-i\bar{\delta}(k))],
\label{Stokes}
\end{eqnarray}
where $\Gamma (z)$ is the complex gamma function. 
Here, we comment briefly on the derivation of Eq.~(\ref{LZ_matrix}) and the Stokes phase $\gamma_{nb}(k)$. 
First, the LZ transfer matrix of Eq.~(\ref{LZ_matrix}) is a slightly modified version of the usual LZ transfer matrix derived in \cite{Nori} 
because $\Gamma_{\pm} (k)$ has an additive phase factor $e^{-i(\pm \tilde{\rho}(k))}$ in the diagonal elements compared with the usual LZ transfer matrix \cite{Nori}.
The phase factor comes from the factor $e^{\pm i\tilde{\rho}(k)}$ in the off-diagonal part in Eq.~(\ref{LZH4}). 
The usual LZ transfer matrix is obtained when the off-diagonal part of Eq.~(\ref{LZH4}) is $\sqrt{A^{2}(k)+B^{2}(k)}$ \cite{Nori}.
In ~\cite{Lim,Lim2}, the LZ transfer matrix of Eq.~(\ref{LZ_matrix}) is derived from Eq.~(\ref{LZH4}). 
Its derivation process is as follows: \\
\nin
(i) Although the factor $e^{\pm i\tilde{\rho}(k)}$ exists in the off-diagonal part in Eq.~(\ref{LZH4}), 
the Schr\"{o}dinger equation of the linearized Hamiltonian in Eq.~(\ref{LZH4}) provides the Weber equation \cite{Zener}. 
Accordingly, the same procedure as the derivation given in \cite{Nori} can be applied.\\
\nin
(ii) The Weber equation can be solved asymptotically. The solution is given by the Weber function \cite{Zener,Nori}.
Thus, the solution gives the probability amplitudes for the upper and lower band states. 
Then, the phase part of the Weber function is the origin of the part of the Stokes phase \cite{Nori,Kayanuma,Kayanuma2}.\\
\nin
(iii) Then, the Weber function solution is substituted into the Schr\"{o}dinger equation for the linearized RM Hamiltonian in Eq.~(\ref{LZH4}), 
and a slightly modified expression of Eq.~(A.12) in \cite{Nori} can be obtained, which includes the $e^{-i(\pm \tilde{\rho}(k))}$ factor.\\
\nin
(iv) Using the modified expression of Eq.~(A.12) in \cite{Nori} and following the same procedure in appendix A in \cite{Nori}. 
the LZ matrix of Eq.~(\ref{LZ_matrix}) can be obtained.

Here, $\Gamma_{+} (k)$ and $\Gamma_{-} (k)$ act as a transfer matrix between the lower and upper bands at $t_{1}$ and $t_{2}$, respectively.
In the matrix $\Gamma_{\pm} (k)$, the diagonal terms are lower-to-lower and upper-to-upper state transitions, 
and the off-diagonal terms are a lower-to-upper state transition and vice versa. 
As explained in detail below, the adiabatic-impulse approximation is employed. 
Therefore, the matrix $\Gamma_{\pm}(k)$ acts on extremely narrow time intervals, $t_{1}-0\leq t\leq t_{1}+0$ and $t_{2}-0 \leq t\leq t_{2}+0$. 
Accordingly, we ignore the dynamical phase accumulated in such a narrow time regime.  

In addition, we should comment on the approach of the adiabatic perturbation theory (APT) carried out in \cite{Privitera}.
Both the ATP and LZ transition approaches focus on the small $\Omega$ regime. 
In the former approach, the Floquet lowest energy population after a single pumping cycle has a $\mathcal{O}(\Omega^{3})$ error, and 
thus for a large $\Omega$ regime, the former approach is broken. 
In the later approach, as long as the double avoided crossings in the band structure are created by tuning the parameters 
and the largest gap regime in the band is sufficiently larger than $\Omega$, a large transfer to the excitation band at the avoided crossing can be handled, 
and using the LZ transfer matrix in Eq,~(\ref{LZ_matrix}) it is possible to incorporate the influence of the interband coherence.


\section{Time evolution in adiabatic-impulse approximation}
\label{TEA}
\begin{figure}[t]
\begin{center} 
\includegraphics[width=7.5cm]{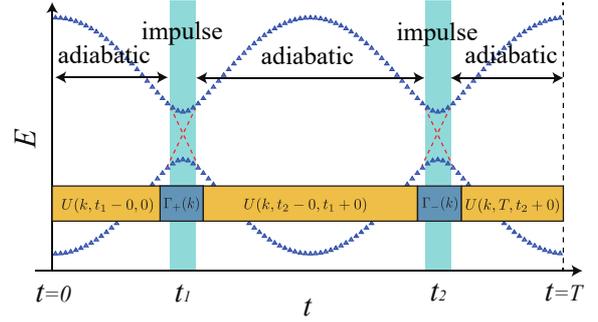}
\end{center} 
\caption{Schematic figure of the adiabatic-impulse approximation and the Landau--Zener transition.
The blue shaded area represents an impulse regime, which is an extremely narrow time interval.
In the adiabatic regime, the transition between the upper and lower bands is negligible. 
In the blue shaded regime, the system is described using the linearized Hamiltonian $\tilde{h}_{RM}(k,\delta t)$. }
\label{Adiabatic-impluse}
\end{figure}
The adiabatic-impulse approximation is applied to the band structure, as shown in Fig.~\ref{RMband} (a).  
Here, we assume that the avoided crossing regime for the target parameter regime is narrow. 
In the other time regimes, the system is under adiabatic conditions.
A schematic figure of the adiabatic-impulse approximation is shown in Fig.~\ref{Adiabatic-impluse}. 
Under these conditions, the time evolution of the wave function expanded by the instantaneous eigenstates is considered, with a focus on the change in the lower band occupation.  
First, we prepare the wave function constituted by a linear combination of the periodic functions of the instantaneous lower and upper bands,
\begin{eqnarray}
|\Psi(k,t)\rangle=c_{1}(k,t)|u_{1}(k,t)\rangle+c_{2}(k,t)|u_{2}(k,t)\rangle, 
\end{eqnarray}
where $c_{1(2)}(k,t)\in {\bf C}$ is the coefficient of the instantaneous eigenstate of the lower (upper) band, and $|u_{1(2)}(k,t)\rangle$ is determined by the Bloch theorem at time $t$. 
The time evolution of $c_{1(2)}(k,t)$ can then be calculated 
from the adiabatic-impulse approximation. 
In the adiabatic regime, the time evolution is obtained 
by considering the following unitary operator:
\begin{eqnarray}
U (k,t,t') &=&
\left[ 
\begin{array}{cc} 
e^{-i\int^{t}_{t'} E_{+}(k,t'')dt''} & 0  \\ 
0 & e^{-i\int^{t}_{t'} E_{-}(k,t'')dt''} \\ 
\end{array} 
\right].\nonumber\\
\label{Unitary_operator}
\end{eqnarray}
The operator $U$ acts on the coefficient vector $(c_{2},c_{1})^{t}$ and gives the adiabatic time evolution from $t'$ to $t$.
By contrast, around the avoided crossing corresponding to the impulse regime, 
the time evolution of $c_{1(2)}(k,t)$ can be obtained by applying the LZ transition matrix $\Gamma_{\pm}(k)$. 
It should be noted that the width of the impulse regime is assumed as a single point. 
Even under such an assumption, the adiabatic-impulse approximation is believed to give fairly correct results, as mentioned in \cite{Nori}.
Under this situation, the coefficient $c_{1(2)}(k,T)$ can be connected to $c_{1(2)}(k,0)$ by applying the unitary
operator $U$ and the LZ transition matrix $\Gamma_{\pm}(k)$. By introducing the coefficient vector defined by ${\bf c}(k,t)=(c_{2}(k,t),c_{1}(k,t))^{t}$, 
the one-cycle time evolution can be written in the following form:
\begin{eqnarray}
{\bf c}(k,T)& =& U(k,T,t_{2}+0)\Gamma_{+}(k) U(k,t_{2}-0,t_{1}+0) \nonumber\\
&\times&\Gamma_{-}(k) U(k,t_{1}-0,0){\bf c}(k,0).
\label{TEc1}
\end{eqnarray}
From this relation, the time evolution of the wave function $|\Psi (k,t)\rangle$ can be obtained. The time evolution includes the interband transition effect. 
The interband transition is a non-adiabatic effect.
In this work, we put $c_{1}(k,0)=1$ as an initial state.
For the time evolution described by Eq.~(\ref{TEc1}), in this study we focus only on the dynamics of the lower band, i.e., the dynamics of $c_{1}(k,t)$, is of key interest. 
The contribution of the lower-to-upper band after one pumping cycle may be regarded as a dissipation from the lower band state \cite{Oka}. 
In this work, however, we do not focus on this contribution. 

From Eq.~(\ref{TEc1}), the lower band population denoted by $|c_{1}(k,T)|^2$ is given by
\begin{eqnarray}
|c_{1}(k,T)|^{2}&=&1-2p_{LZ}(k)+2(1-p_{LZ}(k))p_{LZ}(k) \nonumber\\
&\times&\cos \biggl[ \int^{t_{2}}_{t_{1}}[E_{+}(k,t')-E_{-}(k,t')] dt'-2\gamma_{nd}(k)\biggr].\nonumber\\
\label{c1kT}
\end{eqnarray}
\noindent Here, it should be noted that in Eq.~(\ref{c1kT}) the coefficient $c_{1}(k,t)$ picks up the Stokes phase $\gamma_{nb}(k)$ twice because the LZ transition matrix $\Gamma(k)$ acts twice along the time evolution. 
Under a classical assumption, the lower band population may be $|c_{1}(k,T)|^{2}=1-p_{LZ}(k)+p^{2}_{LZ}(k)$; however, Eq.~(\ref{c1kT}) includes the cosine factor, 
whose phase factor is determined by the information regarding the instantaneous energy spectrum of both the lower and upper bands. 
Thus, the cosine factor can be regarded as the interference effect of the upper band. 

In addition, we mention the large $T$ limit for Eq.~(\ref{c1kT}). We then obtain $|c_{1}(k,T)|^{2}\to 1-2[1-\cos(\beta)]p_{LZ}(k)$, 
where $\beta$ is the phase of the cosine part of the RHS in Eq.~(\ref{c1kT}). 
Because $p_{LZ}$ decays exponentially with the increase of $T$, $|c_{1}(k,T)|^{2}$ tends to decay exponentially with the increase of $T$.
It is also interesting to compare Eq.~(\ref{c1kT}) with the result in \cite{Privitera}. 
In \cite{Privitera}, a Floquet analysis shows that the lowest order deviation from unity in the nonadiabatic population transfer is proportional to $\Omega^2$. 
Although it is difficult to verify the complete relationship between our result of Eq.~(\ref{c1kT}) and the $\Omega^2$ decay behavior, 
as in \cite{Privitera}, if the lowest energy Floquet state is prepared as the initial state through a suitable smooth switch-on of the driving, 
exponentially small corrections are recovered. Accordingly, Eq.~(\ref{c1kT}) in the large limit of $T$ is expected to exhibit 
similar behavior as the result obtained by the Floquet analysis in a suitable smooth switch-on driving case.


\section{Non-adiabatic extension of the Zak phase}
\label{NAEZ}
Using Eq.~(\ref{c1kT}), a non-adiabatic extension of the Zak phase can be formulated. 
The Zak phase is known to correspond to the electric charge polarization in a strongly correlated electron system \cite{Resta,Vanderbilt,Marzari}. 
To formulate the extension form, by first using the lower band sector of $|\Psi(k,t)\rangle $, 
we construct the lower band Wannier function $|W(t)\rangle$ \cite{Asboth} as follows:  
\begin{eqnarray}
|W(t)\rangle &=& \frac{1}{\sqrt{N}}\sum^{N}_{m=1}e^{imk}|m\rangle \otimes c_{1}(k,t)|u_{1}(k,t)\rangle \nonumber\\
&=&  \int^{\pi}_{-\pi}\frac{dk}{2\pi}|k\rangle \otimes c_{1}(k,t)|u_{1}(k,t)\rangle, \label{Wani}
\end{eqnarray}
where $m$ is a lattice site, $|m\rangle$ is the state in which a particle is localized at site $m$, and $N$ is the total number of lattice sites. 
Accordingly, from $|W(t)\rangle$, the CM of the lower band Wannier function is given as $\langle W(t)|\hat{x}|W(t)\rangle$, 
where $\hat{x}$ is the position operator of the particle as viewed from a continuous space.
In general, the CM is known to correspond to the Zak phase \cite{Asboth,Vanderbilt,Marzari}. 
Hereafter, the CM $\langle W(t)|\hat{x}|W(t)\rangle$ is denoted by $P(t)$.
If $c_{1}(k,0)=1$, i.e., the initial state at $t=0$ is in the lower band state and we use $c_{1}(k,T)$ obtained from Eq.~(\ref{TEc1}), 
the CM at $t=T$, $P(T)$ can be calculated as follows:
\begin{eqnarray}
P(T)&=&\frac{i}{2\pi}\int^{\pi}_{-\pi}dk\: \biggl[c^{*}_{1}(k,T)\partial_{k}c_{1}(k,T)\nonumber\\
&&+|c_{1}(k,T)|^{2}\langle u_{1}(k,T)|\partial_{k}u_{1}(k,T)\rangle \biggr],
\label{Pt}
\end{eqnarray}
where the first term in the integrant in the LHS vanishes 
because $c^{*}_{1}(k,T)$ and $c_{1}(k,T)$ are symmetric for $k$ because $E_{\pm}(k,t)$, $\bar{\delta}(k)$, and $p_{LZ}(k)$ are symmetric for all $k$, 
i.e, $c^{(*)}_{1}(k,T)=c^{(*)}_{1}(-k,T)$. Therefore, $P(T)$ is determined by only the lower band population $|c_{1}(k,T)|^{2}$ after a single pumping cycle.
In addition, because in an adiabatic limit $T\to \infty$, $|c_{1}(k,T)|\to 1$, the representation $P(T)$ is smoothly connected to the usual (adiabatic) Zak phase form \cite{Asboth}. 
In this sense, Eq.~(\ref{Pt}) can be regarded as a non-adiabatic extension form of the Zak phase \cite{nonZak}.

Furthermore, by using Eq.~(\ref{Pt}), we can directly write the displacement of $P(t)$ from $t=0$ to $t=T$ as follows: 
\begin{eqnarray}
&&P(T)-P(0) \equiv \Delta P(T) \equiv \Delta P_{0}(T)+\delta P(T),\label{DPT}\\
&&\Delta P_{0}(T)\equiv\frac{i}{2\pi}\int^{\pi}_{-\pi} dk\: \biggl[\langle u_{1}(k,T)|\partial_{k} u_{1}(k,T)\rangle\nonumber\\
&&\:\:\:\:\:\:\:\: -\langle u_{1}(k,0)|\partial_{k} u_{1}(k,0)\rangle\biggr],\label{DPT2}\\
&&\delta P(T) \equiv \frac{i}{2\pi}\int^{\pi}_{-\pi} dk\: \gamma_{d}(k) \langle u_{1}(k,T)|\partial_{k} u_{1}(k,T)\rangle,\\
&&\gamma_{d}(k) \equiv |c_{1}(k,T)|^{2}-1,
\end{eqnarray}
where $|c_{1}(k,T)|^{2}$ is given by Eq.~(\ref{c1kT}), $\Delta P_{0}(T)$ represents the adiabatic portion of the displacement $\Delta P(T)$, and 
$\gamma_{d}(k)$ represents the deviation from the full population of the lower band for each $k$.  

Here, the meaning of the total deviation $\Delta P(T)$ should be further discussed: $\Delta P(T)$ is the total shift of the CM of the lower band Wannier function after a single pumping cycle. 
The total deviation $\Delta P(T)$ corresponds to the lower band pumped charge, 
and not the total pumped charge of the system, which is generated by both the lower and upper band contributions. Here, to distinguish them we denote the two charges by $Q_{L}$ and $Q_{sys}$, respectively. 
In an adiabatic limit, the lower band pumped charge $Q_{L}$ 
corresponds to the total pumped charge $Q_{sys}$ of the system because the upper band contribution is negligible. 
This situation corresponds to the usual topological charge pumping \cite{Asboth}.
Therefore, we clearly define the lower band pumped charge $Q_{L}$ as 
\begin{eqnarray}
Q_{L} \equiv \Delta P(T).
\end{eqnarray}
Here, if we assume the adiabatic limit $T\to \infty$, $Q_{L}$ is as follows: 
\begin{eqnarray}
Q_{L} = \Delta P_{0}(T) = C_{N}, \label{CNP}
\end{eqnarray}
where we use the fact that $\Delta P_{0}(T)$ is regarded as the lower band Chern number $C_{N}$ \cite{Asboth}. Thus, in adiabatic situation, 
because $C_{N}$ is known to take an integer value \cite{Thouless,TKNN}, 
the lower band pumped charge current $Q_{L}$ takes an integer value, 
i.e., the topological charge pumping is recovered. However, in non-adiabatic situation, $Q_{L}$ does not take an integer value. 
This indicates the breakdown of the quantization of the topological charge pumping owing to the decay of the lower band population $|c_{1}(k,T)|^{2}$.

\section{Estimation of the LZ formulation}
\label{ELZ}
\begin{figure}[t]
\begin{center} 
\includegraphics[width=6cm]{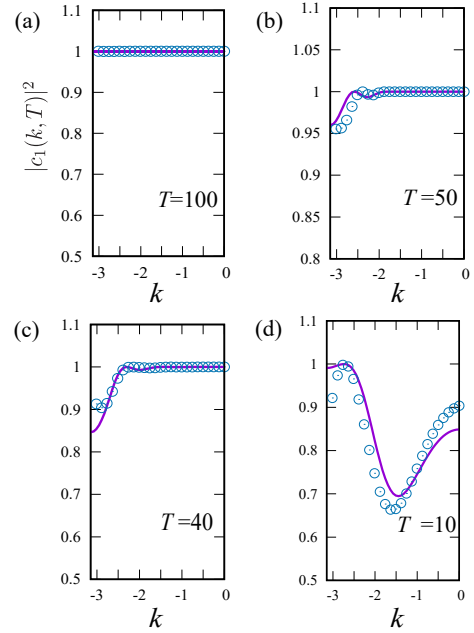}
\end{center} 
\caption{Lower band population $|c_{1}(k,T)|^{2}$ for $T=100 $(a), $50$(b), $40$(c), and $10$(d). 
The solid line shows the analytical result of Eq.~(\ref{c1kT}). The blue circle shows the numerical result.}
\label{c1pop}
\end{figure}
In this section, we evaluate the adiabatic impulse approximation and the LZ formulation with the help of a numerical simulation.
In particular, we estimate the lower band population $|c_{1}(k,T)|^{2}$, and 
demonstrate how the LZ result of Eq.~(\ref{c1kT}) captures the weak non-adiabatic dynamics in the RM model.
To this end, we numerically calculate the dynamics of the RM model obeying the Schr\"{o}dinger equation. 
We then use the momentum representation of the Schr\"{o}dinger equation, 
\begin{eqnarray}
i\frac{d}{dt}|\Psi(k,t)\rangle = \hat{h}_{\rm RM}(k,t)|\Psi(k,t)\rangle, \label{Seq}
\end{eqnarray}
and use the spin up and down bases, $|\Psi(k,t)\rangle=a_{1}(k,t)|\uparrow\rangle+a_{2}(k,t)|\downarrow\rangle$, 
where $a_{1(2)}(k,t) \in {\rm {\bf C}}$ and $\hat{\sigma}_{z}|\uparrow (\downarrow)\rangle=+1(-1)|\uparrow (\downarrow)\rangle$. 
In solving Eq.~(\ref{Seq}), we employ a fourth-order Runge-Kutta method. 
To obtain the lower band population at $t=T$ after one pumping cycle, 
we employ a gauge fixed exact solution of the instantaneous eigenvector of the lower band in the RM model. 
This is given by 
\begin{eqnarray}
|u^{ex}_{1}(k,t)\rangle =
\left(
\begin{array}{c}
\cos[\phi(k,t)/2]\\
-e^{-i\theta(k,t)}\sin[\phi(k,t)/2]
\end{array}
\right), \label{exactu1}
\end{eqnarray}
where the parameters $\phi(k,t)$ and $\theta(k,t)$ are determined by $J_{1}+J_{2}e^{ik}=[\tan\phi(k,t)]e^{i\theta(k,t)}$.
By solving Eq.~(\ref{Seq}) for each $k$ and using the exact solution $|u^{ex}_{1}(k,t)\rangle$,  
we can obtain the numerical result of the lower band population at $t=T$ as $|\langle u^{ex}_{1}(k,T)|\Psi(k,T)\rangle |^{2}$.

Let us estimate the LZ analytical form of Eq.~(\ref{c1kT}). 
We calculate both the LZ analytical lower band population \cite{anagamma} and the numerical one obtained by solving Eq.~(\ref{Seq}).
In this study, we vary the sweeping speed by varying $T$ for the one pumping cycle from $T=10$ ([$\hbar/\Delta_0$]) to $T=100$.
Here, $T$ is connected to the driving frequency as $\Omega=2\pi/T$. 
In our target parameter set in the RM model, the minimum band gap at the two avoided crossings is $4\delta_{0}=0.8$ (we take $\Delta_{0}$ as the energy unit).
Fig. \ref{c1pop} shows the analytical and numerical results for $T=100$, $50$ $40$, and $10$ in $-\pi\leq  k \leq 0$. 
The $T=100$ case in Fig.~\ref{c1pop} (a) is assumed to be adequately adiabatic 
because the driving frequency is quite small for the minimum band gap, i.e., $\Omega \ll 4\delta_{0}$. 
The result therefore indicates that both the analytical and numerical lower band population $|c_{1}(k,t)|^{2}$ remains completely in the lower band after one pumping cycle. 
Therefore, our analytical form of Eq.~(\ref{c1kT}) covers the adiabatic dynamics of the lower band population.

Next, see Fig.~\ref{c1pop} (b), (c), and (d), these are for $T=50$, $40$ and $10$. 
In this parameter regime, we observed some decays of the lower band population, where the upper band population is finite at $t=T$. 
For $T=50$ in Fig.~\ref{c1pop}(b), please see the analytical result. 
Although the driving $\Omega$ is small, that is, the situation is fairly adiabatic, 
a decay from $|c_1(k,T)|^{2}=1$ occurs, particularly around $k=-\pi$, because the band gap at $k=-\pi$ is smallest in the first Brillouin Zone, 
and the analytical result is almost consistent with the numerical calculation, which is shown with the blue circles in Fig.~\ref{c1pop}(b).

Furthermore, let us focus on a faster driving case. As shown in Fig.~\ref{c1pop} (c), the analytical result for $T=40$ seems to indicate 
that the degree of decay around $k=-\pi$ is larger than that in the case of $T=50$. 
Although the numerical result seems to almost capture the behavior of the analytical result, slightly different values in the analytical result $|c_{1}(k,T)|^{2}$ are shown at near $k=-\pi$. 
We expect that this difference originates from the adaptability of the adiabatic-impulse approximation and the LZ formalism. 
More concretely, as noted in \cite{Nori}, the validity of the LZ transition method 
is determined based on an inequality condition $4[d^{2}_{x}(k,t_{1(2)})+d^{2}_{y}(k,t_{1(2)})+\delta^{2}_{0}]\gg \Omega^{2}$, where the LHS is proportional to the band gap.
We estimate this condition for our target case. At near $k=-\pi$, the band gap is small, and thus the LHS tends to be small. 
However, the RHS tends to be large for a small $T$. Therefore, under this situation, the degree of the inequality condition is weak as approaching the minimum band gap around $k=-\pi$ \cite{detaildecay}.
As a result, as shown in Fig.~\ref{c1pop} (c), we expect 
that a deviation of the analytical result from the numerical result tends to appear at near $k=-\pi$ for a small $T$. 
This result indicates the limitation of the adiabatic-impulse approximation and the LZ transition method.
Such a deviation tendency also appears for a further fast driving case. 
Please see the $T=10$ case in Fig.~\ref{c1pop} (d). Although at a glance, the analytical result of the lower band population is almost consistent with the numerical result,
 a large deviation occurs between the analytical result and numerical one around $k=-\pi$, 
and also in this case we find a further deviation at near $k=0$. 
We expect that this finding may also originate from the same reason with around $k=-\pi$. 
In addition, we should comment the maximum decay regime for $T=10$. 
Seen from the band structure and band gap tendency, as shown in Fig.~\ref{RMband}, 
the maximum decay is intuitively expected to occur around the minimum band gap regime $k=\pm \pi$. 
As shown in our analytical and numerical results, this intuition is true for $T=50$ and $40$, as shown in Fig.~\ref{c1pop} (b) and (c), 
but not for $T=10$. We expect that this difference originates from the interference effect of the upper band, 
which is determined by the difference between the lower and upper band instantaneous energy spectrums and the Stokes phase as Eq.~(\ref{c1kT}). 

Furthermore, it is interesting to compare $\Delta P(T)$, which was introduced in the previous section, with the total pumped charge $Q_{sys}$ of the system.
In particular, we compare the analytical $\Delta P(T)$ with the numerical $Q_{sys}$ obtained by solving Eq.~(\ref{Seq}).
Numerically, the total charge current of the system can be calculated by \cite{Asboth,Lim3} 
\begin{eqnarray}
J(t)=\int^{\pi}_{-\pi}\frac{dk}{2\pi}\langle\Psi(k,t)|\frac{\partial\hat{h}_{\rm RM}(k,t)}{\partial k}|\Psi(k,t)\rangle \label{Current},
\end{eqnarray}
Using the current $J(t)$, the total system pumped charge is given by
\begin{eqnarray}
Q_{sys}(t)=\int^{t}_{0}dt'\: J(t') \label{Charge_pumped_Current}.
\end{eqnarray}
\begin{figure}[t]
\begin{center} 
\includegraphics[width=8.5cm]{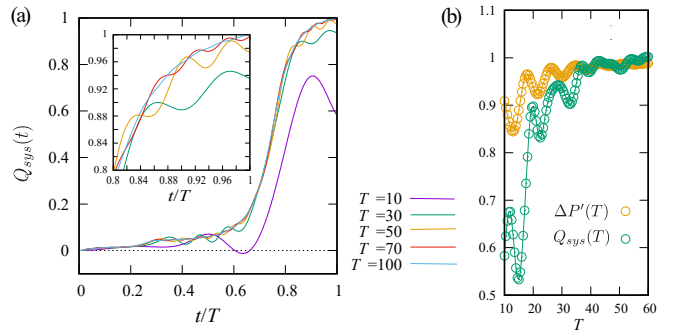}
\end{center} 
\caption{(a) $T$-dependence of the total pumped charge of the system. 
(b) The decay behavior of $\Delta P'(T)$ and $Q_{sys}(T)$ when varying $T$. }
\label{dQf}
\end{figure}

By contrast, it is difficult to directly calculate $\Delta P(T)$ of Eq.~(\ref{DPT}) owing to the gauge dependence. 
To avoid this difficulty, we shift the target time interval of the RM model, $\Omega t\to \Omega \tilde{t}=\Omega (t-T/4+\delta t)$, where $\delta t$ is a positive small displacement. 
Then, without a loss of generality, we can construct the same CM shift $\Delta P'(T)\equiv P'(T)-P'(0)$, where $\tilde{t}=0$ and $\tilde{t}=T$ are extremely close to the inversion symmetric point of the RM model.
Under adiabatic condition, the values of $P'(T)$ and $P'(0)$ are known to be $1/2$ and $-1/2$, respectively \cite{Asboth,Zak}.
In non-adiabatic regime, the finite $\gamma_{d}(k)$ in $P(T)'$ of Eq.~(\ref{DPT}) causes a breakdown of the discrete gauge invariance $P'(T)\to P'(T)+j$ (where $j$ is an arbitrary integer) \cite{Asboth}. 
Thus, under non-adiabatic situation, while $P'(0)$ remains in $-1/2$ by using the exact solution of the instantaneous eigenvector of the lower band in Eq.~(\ref{exactu1}), 
$P'(T)$ is expected to deviate from the value of $1/2$.
Here, to estimate the value of $\Delta P(T)$ qualitatively, we assume that the equivalence $\Delta P(T)=\Delta P'(T)$ is almost consistent. We calculate $\Delta P'(T)$ 
using the exact solution of the instantaneous eigenvector of the lower band in Eq.~(\ref{exactu1}). 

To begin with, in Fig.~\ref{dQf}(a) we plot the time-dependence of $Q_{sys}(t)$ for various sweeping speeds. For $T=100$ and $70$, $Q_{sys}(T)=1$, 
whereas for $T=50$, $Q_{sys}(T)$ does not reach unity, which clearly represents a breakdown of the adiabatic condition and the quantization of the topological charge pumping.
Figure~\ref{dQf}(b) shows the $T$-dependence of $\Delta P'(T)$ and $Q_{sys}(T)$. 
Both cases clearly show a breakdown of the adiabatic condition and deviate from unity when decreasing $T$. 
Interestingly, the decay behavior represents an oscillatory damping. This is expected to be caused by the interference term in Eq.~(\ref{c1kT}). 
In addition, the result displays a reasonable behavior for a small $T$ regime, $\Delta P'(T)>Q_{sys}(T)$. 
This is because $\Delta P'(T)$ includes only the lower band CM shift, 
whereas $Q_{sys}(T)$ includes the current contribution of both the lower and upper bands, 
where the current contribution of the upper band is inverse to that of the lower band, which originates from the upper band Chern number $C_{N}=-1$ \cite{Asboth}.
Remarkably, figure.~\ref{dQf}(b) indicates that, in the LZ formulation, a non-adiabatic breakdown of $\Delta P'(T)$ and $Q_{sys}(T)$ occurs 
from a slow sweeping speed to a certain extent compared to the inverse minimum band gap at the avoided crossing.

\section{About experiment}
\label{EXPLZ}
Here, we mention the verification of our results for a cold atom optical lattice experiment.
Measuring the CM is not too difficult because an experimental method has previously been established, e.g., a band mapping method \cite{Lohse,Nakajima}. 
The RM model has already been implemented in an optical super lattice system \cite{Lohse,Atala} 
and a continuous RM model has also been developed \cite{Nakajima}. 
These experimental systems can reach our considered parameter regime in terms of $J_{0}$,$\delta_{0}$ and $\Delta_{0}$ in the RM model.
However, some experimental limitations still remain.  
For example, a perfect full occupation of the lower band state has not been realized \cite{Nakajima} 
owing to a finite temperature effect, and a harmonic trapping potential in the experimental systems also breaks the translational symmetry of the system.  
These obstacles must be overcome before highly accurate measurements of both the lower band population 
after one pumping cycle and the CM shift of the lower band Wannier function are carried out 
because the deviation from the quantization value in our estimation described in this study is small, i.e., at most $5-10\%$ in a weak non-adiabatic regime. 

In addition, throughout this work, we focused on the weak non-adiabatic effects, i.e., a small correction for the lower band population.
Our estimation for such a small correction can also provide a deeper understanding and a key to a high precision control of quantum devices such as superconducting qubits \cite{Nori,Krantz}.

\section{Conclusion}
\label{Conq}
A weak non-adiabatic effect for the Zak phase and the topological charge pumping in the RM model has been discussed. 
The dynamics of the lower band state has been formulated by applying an adiabatic-impulse approximation and using the LZ transition method. 
We have derived the lower band population after one pumping cycle. The formula of Eq.~(\ref{c1kT}) includes the interference effect from the upper band. 
From the lower band population after one pumping cycle, 
we have obtained a LZ analytical formula describing a non-adiabatic extension of the Zak phase, which corresponds to the total CM shift of the lower band Wannier function. 
We then estimated the validity of the LZ analytical method. In particular the analytical lower band population has been estimated 
through a comparison with the results of a dynamic numerical simulation. 
The analytical lower band population of Eq.~(\ref{c1kT}) almost captures the non-adiabatic dynamics of the lower band population in the system for the weak adiabatic regime.
Furthermore, the breakdown of the quantization of the total pump charge of the system has been numerically evaluated for various sweeping speeds and 
compared with the CM shift of the lower band Wannier function obtained using the LZ analytical method. 
The results indicate that the decay behavior depending on the sweeping speed exhibits some oscillating behavior, which may originate from the interference factor in Eq.~(\ref{c1kT}).  
In addition, we found that the breakdown of the quantization of the topological charge pumping 
starts earlier than the qualitative starting point characterized by the minimum band gap in the RM model.

If cold atom optical lattice experimental system is further cooled 
and achieves the full occupancy for the lower band in the experimental RM model, our finding can be measured. 

The general idea and prescription used to derive Eqs.~(\ref{c1kT}), (\ref{DPT}), and (\ref{DPT2}) 
are effective in investigating the non-adiabatic effects in wider topological models. \\

Y. K. acknowledges the support of a Grant-in-Aid for JSPS
Fellows (No. 17J00486).

\section*{Author contribution statement}
Yoshihito Kuno developed the idea of this article, performed
all calculations and wrote the manuscript.

\bibliographystyle{spphys-tv}
\bibliography{../00Bibtex/rareregions}

\begin{thebibliography}{10}
\bibitem{nonZak}
The decay of the lower band population appears as an amplitude reduction of $|u_{1}(k,T)\rangle$.
The amplitude reduction can be represented as an imaginary phase factor for $|u_{1}(k,T)\rangle$.
Therefore we can reagred Eq.~(\ref{Pt}) as an extended geomentrical object. 
Accordingly, we call Eq.~(\ref{Pt}) a nonadiabatic extension of the Zak phase.
\bibitem{anagamma}
In calculating the complex gamma function $\Gamma(1-i\bar{\delta}(k))$ in the Stokes phase in Eq.~(\ref{c1kT}), 
we employed the Taylor expansion of inverse gamma function.
\bibitem{detaildecay}
For example, when we consider $T=40$ case, in our RM parameter, 
the value of the LHS in the inequality condition are given by $5.28$ for $k=-\pi$ and $8$ for $k=0$, and the value of the RHS is $\Omega^{2}=0.024$.
Therefore, the inequality condition at $k=-\pi$ is weaker than at $k=0$.
\end{thebibliography}

\end{document}